\documentclass[conference]{IEEEtran}
\IEEEoverridecommandlockouts
\usepackage{cite}
\usepackage{amsmath,amssymb,amsfonts}
\usepackage{algorithmic}
\usepackage{graphicx}
\usepackage{textcomp}
\usepackage{xcolor}
\usepackage{balance}
\usepackage{booktabs}
\usepackage{hyperref}
\usepackage[symbol]{footmisc}
\usepackage[]{collab}

\newcommand\name{\textsc{TraceMesh}\xspace}

\newcommand{\ie}{{\em i.e.},\xspace}
\newcommand{\eg}{{\em e.g.},\xspace}
\collabAuthor{zb}{blue}{Zhuangbin Chen}
\definecolor{ballblue}{rgb}{0.13, 0.67, 0.8}
\collabAuthor{zh}{ballblue}{Zhihan Jiang}

\def\BibTeX{{\rm B\kern-.05em{\sc i\kern-.025em b}\kern-.08em
    T\kern-.1667em\lower.7ex\hbox{E}\kern-.125emX}}
\begin{document}

\title{\name: Scalable and Streaming Sampling for Distributed Traces}



\author{\IEEEauthorblockN{Zhuangbin Chen$^{\mathsection}$, Zhihan Jiang$^{\ddag}$, Yuxin Su$^{\mathsection}$, Michael R. Lyu$^{\ddag}$, Zibin Zheng$^{\mathsection\P}$\thanks{$^{\P}$Corresponding author.}}
\IEEEauthorblockA{$^{\mathsection}$Sun Yat-sen University, Zhuhai, China, \{chenzhb36, suyx35, zhzibin\}@mail.sysu.edu.cn \\
$^{\ddag}$The Chinese University of Hong Kong, Hong Kong, China, \{zhjiang22, lyu\}@cse.cuhk.edu.hk}}


\maketitle

\begin{abstract}
Distributed tracing serves as a fundamental element in the monitoring of cloud-based and datacenter systems.
It provides visibility into the full lifecycle of a request or operation across multiple services, which is essential for understanding system dependencies and performance bottlenecks.
To mitigate computational and storage overheads, most tracing frameworks adopt a uniform sampling strategy, which inevitably captures overlapping and redundant information.
More advanced methods employ learning-based approaches to bias the sampling toward more informative traces.
However, existing methods fall short of considering the high-dimensional and dynamic nature of trace data, which is essential for the production deployment of trace sampling.
To address these practical challenges, in this paper we present \name, a scalable and streaming sampler for distributed traces.
\name employs Locality-Sensitivity Hashing (LSH) to improve sampling efficiency by projecting traces into a low-dimensional space while preserving their similarity.
In this process, \name accommodates previously unseen trace features in a unified and streamlined way.
Subsequently, \name samples traces through evolving clustering, which dynamically adjusts the sampling decision to avoid over-sampling of recurring traces.
The proposed method is evaluated with trace data collected from both open-source microservice benchmarks and production service systems.
Experimental results demonstrate that \name outperforms state-of-the-art methods by a significant margin in both sampling accuracy and efficiency.
\end{abstract}


\section{Introduction}
\label{sec:intro}

Modern cloud systems have dramatically shifted the paradigm of software architecture by adopting loosely coupled designs for applications and services.
For example, Uber's architecture is composed of several thousands of microservices~\cite{DBLP:conf/sigsoft/HeFLZ0LR023}, and WeChat system hosts more than 3,000 services to manage billions of daily requests~\cite{DBLP:conf/cloud/ZhouCLWSLGOY18}.
While such modularity design brings the benefit of flexibility and scalability, it also necessitates sophisticated monitoring to navigate the inherent complexities of distributed systems.
As such, distributed tracing has rapidly emerged as an essential management tool in cloud systems~\cite{DBLP:journals/tse/ZhouPXSJLD21,DBLP:conf/cloud/LuoXLYXZDH021}.
This is primarily due to its ability to provide a detailed timeline of a request's journey through a system, enabling developers to identify bottlenecks, latency issues, and other performance anomalies.

In cloud service systems, unusual and edge-case system behaviors are rare by definition, such as tail latency.
To maintain high coverage of outlier system events, substantial trace data may be generated in production systems, resulting in significant overhead and costs related to trace generation, collection, and ingestion.
For example, Google is estimated to generate approximately 1,000 TB of raw traces on a daily basis~\cite{sigelman2010dapper}.
State-of-the-art tracing frameworks, such as Jaeger~\cite{jaeger} and Zipkin~\cite{zipkin}, mitigate this overhead by \textit{head-based sampling}, which sets a small sampling rate (\eg 0.1\%~\cite{sigelman2010dapper}) to collect traces.
Since head-based sampling occurs prior to request execution, the sampling decision is made uniformly at random.
Consequently, the sampled traces contain mostly common-case execution paths, with a lot of overlapping and redundant information.
As an alternative, \textit{tail-based sampling} captures traces for all requests, and decides whether to retain a trace after the trace has been generated.
Tail-based sampling schemes allow \textit{biased sampling} to collect more informative and uncommon traces by considering details such as latency and HTTP status code.

To pursue more effective biased sampling, some learning-based approaches have been proposed~\cite{DBLP:conf/sigsoft/HeFLZ0LR023,DBLP:conf/cloud/Las-CasasMGF18,DBLP:conf/cloud/Las-CasasPAM19,DBLP:conf/icws/HuangCYCZ21,DBLP:conf/noms/GiasGSPOC23}.
They employ machine learning techniques to derive a unique feature representation for traces, which is then utilized to automatically distinguish useful traces from normal ones.
However, while progress has been made, some practical challenges remain unaddressed in this field.
In production environments, traces vary significantly in their characteristics due to the complex and dynamic interactions between services.
Therefore, their feature space can be extremely large, particularly when considering both the structural and temporal features~\cite{DBLP:conf/icws/HuangCYCZ21}.
This could potentially lead to the issue known as the \textit{curse of dimensionality}, compromising the performance and scalability of trace sampling analysis.
Moreover, the emergence of new features in online scenarios poses a significant challenge to the model's \textit{adaptability}.
Existing methods~\cite{DBLP:conf/icws/HuangCYCZ21,DBLP:conf/issre/ZhouZPYLLZZD23} propose heuristic rules to periodically eliminate features deemed irrelevant to the current window of trace data.
In addition, they append new dimensions to feature vectors and modify the model structure accordingly to accommodate the unseen features.
These strategies, however, can incur substantial computational overhead and the resulting model may be sub-optimal, especially with the frequent emergence of new features.

To address these practical challenges, in this paper we propose \name.
It aims to sample uncommon traces for distributed systems in a scalable and streaming way, while trying to maintain a low storage budget.
\name leverages the technique of \textit{streaming Locality-Sensitive Hashing (LSH)}~\cite{DBLP:journals/toc/Har-PeledIM12,DBLP:conf/stoc/Charikar02,DBLP:conf/kdd/ManzoorMA16} to enable efficient similarity computation between traces.
This is done by projecting high-dimensional trace data into a low-dimensional space while preserving their similarity.
In this process, new trace features can be seamlessly incorporated without affecting the dimensionality of the input vectors.
To sample uncommon traces, \name groups evolving trace data into meaningful clusters~\cite{DBLP:conf/sdm/CaoEQZ06}, where traces exhibiting significant deviations or variances will be identified and selected.
In this process, \name dynamically adjusts the classification of traces (\eg from uncommon to common) to prevent accumulating redundant information.
Experimental results on trace data collected from two open-source microservice benchmarks and one production cloud system demonstrate that \name can sample uncommon traces more effectively and efficiently than existing methods.

The major contributions of this work are as follows:

\begin{itemize}
    \item We propose \name, a tail-based trace sampler for cloud service systems, which addresses some practical challenges in this field.
    Specifically, \name performs dimension reduction on trace data to mitigate the efficiency issue raised by the high-dimensional nature of traces.
    It can also adapt seamlessly to new trace features that emerge in streaming scenarios, without changing the input dimensionality or model structure.
    The implementation of \name is publicly available\footnote{\url{https://github.com/OpsPAI/TraceMesh}}.
    
    \item We conduct experiments with trace data collected from open-source benchmark microservices as well as production cloud systems.
    The experimental results demonstrate the effectiveness and efficiency of \name over existing baseline methods.
\end{itemize}

The remainder of the paper is organized as follows.
Section~\ref{sec:background} introduces the background of distributed tracing and the problem statement of this work.
Section~\ref{sec:methodology} describes the proposed methodology.
Section~\ref{sec:evaluation} presents the experiments and experimental results.
Section~\ref{sec:related_work} discusses the related work.
Finally, Section~\ref{sec:conclusion} concludes this work.




    



\section{Background}
\label{sec:background}

\subsection{Distributed Traces and Their Sampling}

Distributed tracing provides a detailed end-to-end view of requests as they traverse complex, multi-tier cloud service systems.
Representative open-source distributed tracing tools include Jaeger~\cite{jaeger}, Zipkin~\cite{zipkin}, SkyWalking~\cite{skywalking}, and Lightstep~\cite{lightstep}. 
According to the specifications of OpenTracing~\cite{opentracing}, a trace is a directed acyclic graph including multiple spans which represent the individual units of work done in a distributed system.
Each trace is assigned a unique trace ID upon the initiation of a request, referred to as the \textit{root span}.
This trace ID is then propagated to subsequent child spans, which serves as a crucial identifier for the construction of a complete, cohesive trace.
Each span encapsulates various attributes, including trace ID, span ID, parent span ID, latency, and additional metadata (\eg IP address or service version).
This enables us to understand the performance characteristics, locate problems, and optimize the system.
Distributed tracing tools operate within live production environments, involving trace transmission, processing, and storage, which inevitably causes significant computational and storage overheads.
For example, WeChat could produce dozens of terabytes of trace data daily~\cite{DBLP:conf/icws/HuangCYCZ21}.
Therefore, sampling has emerged as a prevailing approach to reduce these tracing overheads.
To ensure the utility of the captured data, sampling decisions are coherent per request, \ie a trace is either sampled in its entirety, recording the complete end-to-end execution, or not at all~\cite{DBLP:conf/cloud/Las-CasasPAM19}.

The feasibility of trace sampling is based on the fact that cloud service systems operate under normal conditions most of the time, \eg many services guarantee an SLA of over 99.9\%~\cite{awsSLA}.
Thus, useful traces only manifest in a small fraction of requests, which trace sampling aims to capture and persist.
We would like to emphasize that a useful trace is not exclusively one associated with performance issues.
It can also record a normal request execution that triggers a previously unseen service call graph.
The objective of trace sampling is to identify and discard traces that contain repetitive or redundant system execution information.
As mentioned in Section~\ref{sec:intro}, there are two generic strategies for trace sampling, namely head sampling and tail sampling.
Head sampling makes decision at the beginning of a request, which, while useful for curbing overhead, cannot know a priori whether a request will carry interesting information and should be traced.
Such a random decision-making process tends to miss important and minor traces such as tail-latency traces.
In contrast, tail sampling executes after traces have been generated.
It pays the runtime costs of generating and caching trace data, but in return allows more flexible and biased sampling since the execution results (\eg latency, execution graph) become available.

Based on the idea of tail sampling, some learning-based approaches have been proposed.
They exploit machine learning techniques to automatically analyze and predict the significance of a trace, without explicit feature engineering~\cite{DBLP:conf/cloud/Las-CasasPAM19,DBLP:conf/sigsoft/Zhang0ZSYCY22}.
In this process, two types of features are often employed as key indicators of traces' commonness, namely \textit{structural information} (\ie calling path, span depth) and \textit{temporal information} (\ie span duration).
For example, a special input might trigger an unusual service execution path, or early interruptions could result in incomplete traces.
These scenarios can all be characterized by the structural information of a trace.
On the other hand, even if a trace maintains a usual structure, it may not necessarily indicate normal operation, since the request could experience significant latency.
These two types of features collectively provide a comprehensive picture of traces, which facilitates a more accurate trace analysis.

\subsection{Problem Statement}

\begin{figure*}[t]
    \centering
    \includegraphics[width=0.94\linewidth]{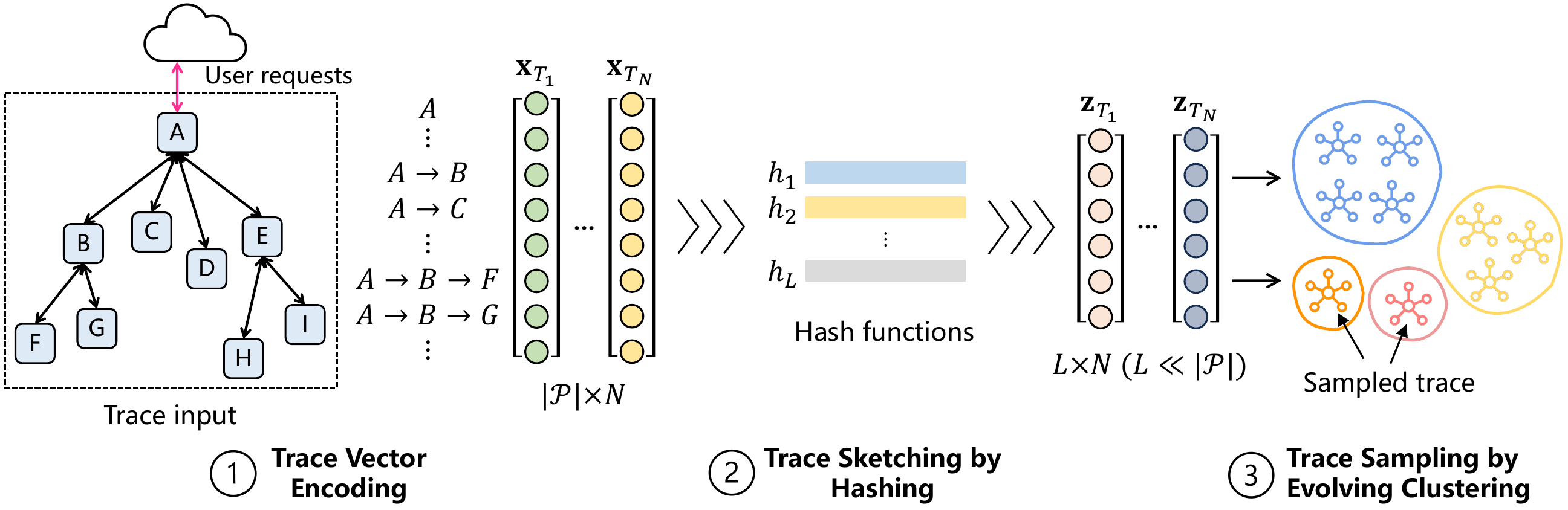}
    \caption{The overall framework of \name}
    \label{fig:framework}
\end{figure*}

The goal of this work is to sample useful traces for cloud service systems to mitigate the computational and storage overheads related to trace processing.
Given a sampling budget (\eg 1\%), we try to identify and capture \textit{uncommon traces} in a streaming scenario, where traces are continuously being generated.
We are interested in two types of traces.
The first type is the unusual traces that record service performance outliers or system failures.
The second type of trace is normal but corresponds to new service operations that haven't been observed in recent periods.
In this process, we try to improve the efficiency of trace sampling by leveraging the idea of dimension reduction.
We also accommodate new features that could emerge in stream data.
These are two essential practical challenges for the deployment of trace sampling techniques in production cloud systems.

\section{Methodology}
\label{sec:methodology}

In this section, we present the design of \name.
The overall framework is illustrated in Figure~\ref{fig:framework}, which consists of three phases, namely, \textit{Trace Vector Encoding}, \textit{Trace Sketching by Hashing}, and \textit{Trace Sampling by Evolving Clustering}.
In the first phase, \name takes graph traces as input, and encodes them as feature vectors by considering both the structural and temporal information.
In the next phase, \name employs streaming LSH to transform the raw trace vectors into sketch vectors, which has a much lower dimensionality and can adapt to unseen features.
In the last phase, \name performs streaming trace sampling through evolving clustering, which adjusts the sampling decision on-the-fly to achieve more accurate results.


\subsection{Trace Vector Encoding}
\label{sec:vector_encoding}

\begin{figure}[t]
    \centering
    \includegraphics[width=0.8\linewidth]{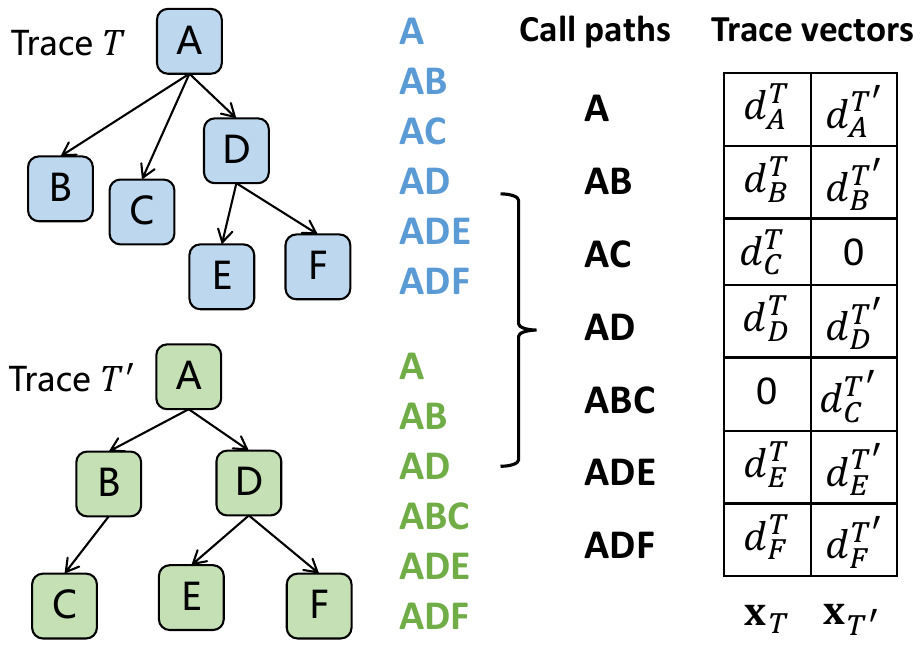}
    \caption{Trace vector encoding}
    \label{fig:trace_encoding}
\end{figure}

Traces are provided in a graph format.
Encoding traces into numerical vectors serves as a prerequisite step for many learning-based methods in trace analysis.
According to Section~\ref{sec:background}, a trace's structural information and temporal information are two essential indicators of its uncommonness.
Therefore, the encoding process should be able to capture and represent these two aspects in an effective and interpretable way.
To this end, we perform trace vector encoding~\cite{DBLP:conf/issre/LiuXOJCZYMZXP20,DBLP:conf/icws/HuangCYCZ21}, as demonstrated in Figure~\ref{fig:trace_encoding}.
A trace records the execution trajectory of a service request.
The initial span of the trace corresponds to the entry point of the request, \ie the root span.
For trace $T$, we start from the first span $A$, and traverse through it in Breadth First Search (BFS) order to visit each of its child spans.
Once a new span (including the first span) is reached, we record the path from the first span to it, and refer to it as a \textit{call path}.
For example, after navigating all spans in trace $T$, we get call paths $A, A\rightarrow B, \dots, A\rightarrow D\rightarrow F$.
Similarly, for trace $T'$, we can extract a new call path $A\rightarrow B\rightarrow C$.
The set of call paths produced after processing all traces is denoted as $\mathcal{P}$, which will be used to construct the entries of the trace vector.
In $\mathcal{P}$, call paths with the same length will be sorted lexicographically for conciseness.
We use duration $d$ associated with the tail span of each call path as the value of the corresponding vector entry. 
For example, the entries $AB$ and $ADE$ of trace $T$ (we omit the symbol $\rightarrow$) have value $d_B^T$ and $d_E^T$, respectively.
In particular, we apply a log transformation, $\lfloor log_{10}d \rfloor$, to normalize the duration.
In reality, minor variations in duration is not practically significant to affect the overall trace pattern.
This allows us to concentrate on the most significant part of the duration, thereby enhancing the stability of trace similarity mining (Section~\ref{sec:vector_hashing}).
Different traces may encompass different sets of call paths. 
For those call paths that are not included within a particular trace (\eg path $ABC$ in trace $T$), their corresponding entries will be assigned a value of 0.
The final trace vector of trace $T$ is represented as $\textbf{x}_T=(d_1, d_2, \dots, d_{|\mathcal{P}|})$, where $|\mathcal{P}|$ is the size of the call path set $\mathcal{P}$.

The trace vectors encoded in this way are capable of distinguishing trace samples with infrequent structural and/or temporal features.
Specifically, if a trace has unique call paths, the resulting vector values will be non-zero, indicating the presence of its uncommon structure.
For two traces that share an identical structure but exhibit significantly different duration statuses, their trace vectors will also differ in the temporal aspect.
The comparison of these aspects can be accomplished by measuring the similarity between trace vectors.
Beyond the path and duration features, \name can also incorporate more meaningful features into the trace vectors to further differentiate uncommon traces.
For example, the request status code or service events can be embedded into the call paths to form more informative trace vectors.

\subsection{Trace Sketching by Hashing}
\label{sec:vector_hashing}

While the vector encoding technique in Section~\ref{sec:vector_encoding} can capture abundant information from a trace, we encounter challenges related to \textit{adaptability} and \textit{dimensionality}.
In production cloud services, the generated traces could encompass a wide variety of span types and exhibit significant length.
Thus, it requires knowing the size of the complete call path set $\mathcal{P}$ to specify the dimension of each trace vector.
With new call paths continuously being formed from the new trace types arriving in stream, the full call path set (and hence its size) always remains undetermined.
Even if the universal call path set is fixed, its size, \ie $|\mathcal{P}|$, can get prohibitively large.
This could potentially lead to the issue known as \textit{the curse of dimensionality}, which could severely impair the performance of downstream trace analysis methods.

To enhance a model's adaptability to previously unseen features, one straightforward way is to adjust the dimensions and then retrain the model.
This is however not scalable, especially when the feature space is dynamically evolving.
Existing methods resort to tree-based models to incorporate new dimensions.
For example, when there is a trace containing paths that never appear before, Sieve~\cite{DBLP:conf/icws/HuangCYCZ21} extends the dimension of other traces by appending $-1$ to their vectors.
The tree structure is subsequently modified by creating a new root.
While this design does accommodate the introduction of new paths, the resulting tree structure may not be optimal.
Therefore, we need a more streamlined approach to handling new paths.
To mitigate the dimensionality problem, prior work~\cite{DBLP:conf/icws/HuangCYCZ21,DBLP:conf/issre/ZhouZPYLLZZD23} has touched upon the idea of dimension reduction.
However, this is achieved by discarding features irrelevant to the current window of trace data, rather than eliminating those that are unimportant.
The discarded features could be those that are introduced in earlier windows to accommodate unseen call paths.
In such a design, certain features can be added and removed for multiple times.
This is not only inefficient, but could damage the model's structure if the number of new dimensions is large or if they manifest frequently.

To address these challenges, we propose to leverage streaming Locality-Sensitive Hashing (LSH)~\cite{DBLP:conf/stoc/Charikar02,DBLP:journals/toc/Har-PeledIM12} for efficient and online trace sampling.
An LSH scheme enables efficient similarity computation by projecting high-dimensional vectors into a low-dimensional space while preserving their similarity.
For instance, by doing so, \textsc{SimHash}~\cite{DBLP:conf/stoc/Charikar02} can quickly measure the cosine similarity between real-valued vectors.
In order to use the \textsc{SimHash} in the streaming setting, Manzoor et al.~\cite{DBLP:conf/kdd/ManzoorMA16} further proposed \textsc{StreamHash}, which is employed in this paper to accommodate unseen call paths.
Details are introduced below.

Given input trace vectors in $\mathbb{R}^{|\mathcal{P}|}$, \textsc{StreamHash} is first instantiated with $L$ projection vectors $\textbf{r}_1, \dots, \textbf{r}_L \in \{+1, -1\}^{|\mathcal{P}|}$.
Each element of $\textbf{r}_l$, $l=1,\dots,L$ is drawn uniformly from $\{+1, -1\}$.
The LSH $h_{\textbf{r}_l}(\textbf{x})$ of an input trace vector $\textbf{x}$ for a given random projection vector $\textbf{r}_l$ is defined as follows:

\begin{equation}
  h_{\textbf{r}_l}(\textbf{x}) =
  \begin{cases}
    +1       & \quad \text{if } \textbf{x} \cdot \textbf{r}_l \geq 0\\
    -1  & \quad \text{if } \textbf{x} \cdot \textbf{r}_l < 0
  \end{cases}
\end{equation}

\noindent That is, $h_{\textbf{r}_l}(\textbf{x})=sign(\textbf{x}\cdot \textbf{r}_l)$.
In particular, $h_{\textbf{r}_l}(\textbf{x})$ possesses the following nice property: the probability (over vectors $\textbf{r}_1, \dots, \textbf{r}_L$) of any pair of input vectors $\textbf{x}_T$ and $\textbf{x}_{T'}$ hashing to the same value is proportional to their cosine similarity:

\begin{equation}
\label{equ:cosine_similarity}
    \textbf{Pr}_{l=1,\dots,L}[h_{\textbf{r}_l}(\textbf{x}_T)=h_{\textbf{r}_l}(\textbf{x}_{T'})]=1-\frac{cos^{-1}(\frac{\textbf{x}_T \cdot \textbf{x}_{T'}}{\parallel \textbf{x}_T \parallel \parallel \textbf{x}_{T'} \parallel})}{\pi}
\end{equation}

Since the computation of similarity between two traces now requires only these hash values, it is feasible to substitute each $|\mathcal{P}|$-dimensional input vector $\textbf{x}$ with a more concise, $L$-dimensional \textit{trace sketch vector} $\textbf{z}$, which encapsulates its LSH values, \ie $\textbf{z}=[h_{\textbf{r}_1}(\textbf{x}),\dots,h_{\textbf{r}_L}(\textbf{x})]$.
In this case, each sketch vector can be succinctly represented using just $L$ bits, where each bit corresponds to a value in $\{+1,-1\}$.
This allows for highly memory-efficient storage algorithms, reducing the complexity of the data structure without compromising its integrity.


The similarity between two input vectors can be subsequently estimated by empirically evaluating the probability presented in Equation~(\ref{equ:cosine_similarity}).
This is done by determining the proportion of aligned hash values when the input vectors are hashed with $L$ random vectors.
As such, the similarity computation between two trace vectors is transformed into a process of quantifying the level of agreement between the hash values.


\begin{equation}
    sim(T,T') \propto \frac{|\{l:\textbf{z}_T(l)=\textbf{z}_{T'}(l)\}|}{L}
\end{equation}

In summary, given a target dimensionality $L \ll |\mathcal{P}|$, each trace \textbf{x} can be represented by a sketch vector \textbf{z} of dimension $L$, allowing us to discard the original $|\mathcal{P}|$-dimensional trace path vectors and compute similarities within this newly-defined vector space.


\begin{figure}
    \centering
    \includegraphics[width=0.98\linewidth]{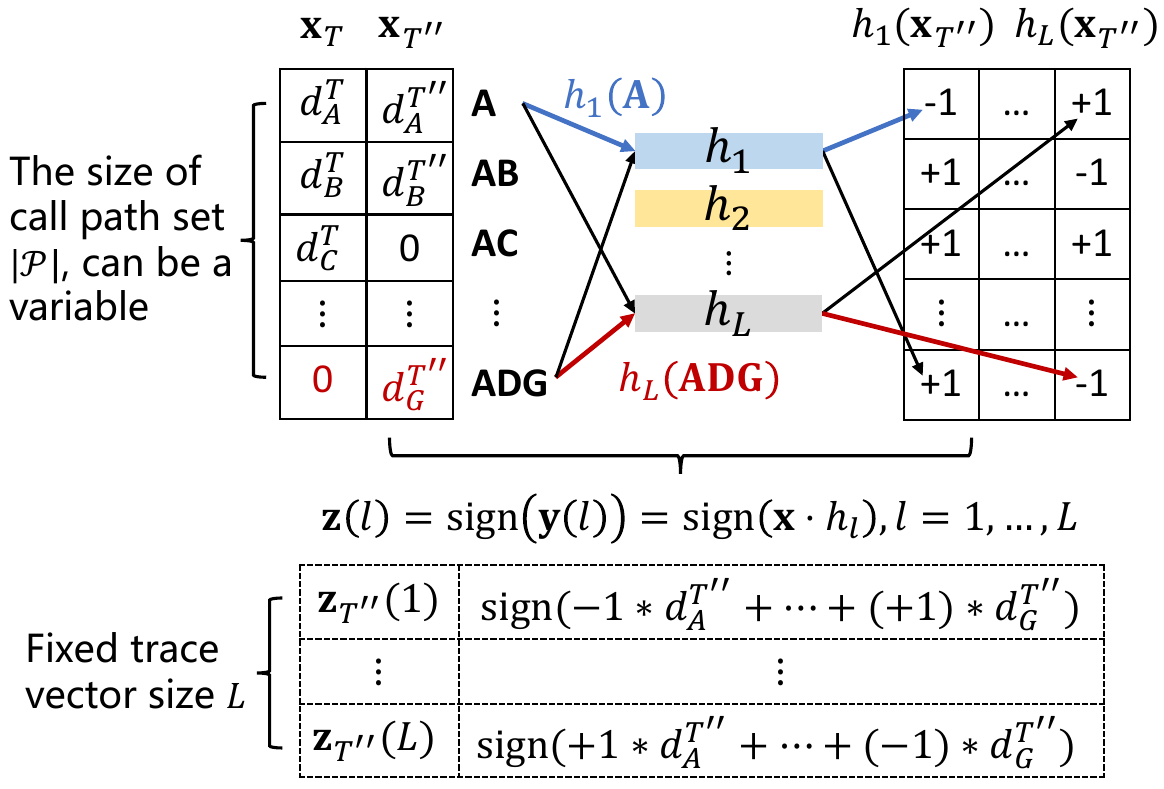}
    \caption{Streaming trace vector encoding}
    \label{fig:trace_sketching}
\end{figure}

While the above operation can effectively mitigate the dimensionality challenge, the adaptability issue remains.
As shown in Figure~\ref{fig:trace_sketching}, suppose a new trace $T''$ comes with a new call path $ADG$ having a duration $d_G^{T''}$.
In this case, many existing studies append a new dimension (in our example, a ``0'') to all the other trace vectors and update the model accordingly.
This could incur significant computational overhead, and the resulting model may be sub-optimal.
To address this problem, instead of using the $L$ projection vectors $\textbf{r}_1, \dots, \textbf{r}_L \in \{+1, -1\}^{|\mathcal{P}|}$ (whose dimensionality is fixed), \textsc{StreamHash} instantiates $L$ hash functions $h_1,\dots, h_L$ picked uniformly at random from a family $\mathcal{H}$ of hash functions, mapping call paths to \{+1,-1\}.
As shown in the top part of Figure~\ref{fig:trace_sketching}, a $h_l\in \mathcal{H}$, $l=1,\dots,L$, is a deterministic function that maps a given call path to either +1 or -1, which replaces the original elements in the projection vectors $\textbf{r}_i$.
To obtain the trace sketch, we first construct the projection vector \textbf{y} of the trace as:

\begin{equation}
    \textbf{y}(l)=\textbf{x}\cdot h_l=\sum_{i=1,\dots,|P|}\textbf{x}(i)h_l(p_i)
\end{equation}

\noindent Then, the $L$-bit trace sketch for each input vector $\textbf{x}$ under the hash functions can be calculated by $\textbf{z}=sign(\textbf{y})$, which can be used to measure the similarity between traces.
In this way, the adaptability issue is also addressed as the trace sketch vectors can be constructed and maintained incrementally, accommodating call paths that are not previously observed.
As a result, we eliminate the need to know the complete call path set $\mathcal{P}$, or to maintain $|\mathcal{P}|$-dimensional random vectors $\textbf{r}_i$ in memory.

In~\cite{DBLP:conf/kdd/ManzoorMA16}, the strongly universal multilinear family~\cite{DBLP:journals/cj/LemireK14} is adopted as the hash function family $\mathcal{H}$ for string data.
As our trace vector encoding also produces string call paths, we employ the same configuration.
In this family, the input call path $p$ is divided into $|p|$ components (\eg spans) as $p=s_1s_2\dots s_{|p|}$.
A hash function $h_l$ is constructed by first choosing $|p|$ random numbers $m_1^{(l)},...,m_{|p|}^{(l)}$, and $p$ is then hashed as follows:

\begin{equation}
    h_l(p)=2\times ((m_1^{(l)}+\sum_{i=2}^{|p|}m_i^{(l)}\times int(s_i))~mod~2)-1
\end{equation}

\noindent where $int(s_i)$ is a function that maps span $s_i$ to a unique integer and $h_l(p)\in \{+1,-1\}$.
We start from one with an incremental growth of one for new span types.
The hash value for a call path of length $|p|$ can be computed in $\Theta(|p|)$ time.

Each hash function is represented by $|p|_{max}$ random numbers, where $|p|_{max}$ denotes the maximum possible length of a call path.
These numbers remain constant per hash function $h_l$, which serve as the parameters.
Thus, the hash functions can deterministically hash a given call path to the same value each time.
In practice, these $L$ hash functions can be generated uniformly at random from this family by creating a matrix of $L\times |p|_{max}$ uniformly random 64-bit integers using a pseudorandom number generator.
While we still need some global information, \ie $|p|_{max}$, this is a much lighter restriction.
Existing work needs to know a priori the concrete type of all call paths.
In cases where a new call path has a length exceeding $|p|_{max}$, we can break it into larger chunks (e.g., two spans constitute one component)~\cite{DBLP:conf/kdd/ManzoorMA16}.


\subsection{Trace Sampling by Evolving Clustering}
\label{sec:trace_sampling}

In real-world systems, most traces share similar and common characteristics.
Trace sampling aims to identify the uncommon traces that exhibit rare patterns and ensure that they are sampled with a higher probability.
In Section~\ref{sec:vector_hashing}, based on the streaming LSH technique, we are able to encode new traces with arbitrary call paths for similarity computation.
Next, \name conducts trace sampling by grouping them into meaningful clusters, where each cluster contains similar traces in both structural and temporal perspectives.
The uncommon traces can then be identified based on the deviations from these established clusters.
In production systems, traces are continuously being generated, \ie streaming data.
During this process, uncommon traces exhibiting unprecedented patterns can emerge.
Thus, our clustering approach should not only recognize the new unusual patters, but also allow the false positives (\ie the new usual patterns) to eventually evolve into normal clusters and stop sampling them.
This strategy ensures a more accurate and efficient trace sampling.

To this end, \name adapts DenStream~\cite{DBLP:conf/sdm/CaoEQZ06} to better fit our task of trace stream clustering.
DenStream is a popular density-based clustering approach for evolving stream data.
Without the assumption on the number of clusters, it can discover clusters with arbitrary shape and handle outliers.
In DenStream, each data point is associated with a weight, which decreases exponentially with time $t$ via a fading function $f(t)=2^{-\lambda \cdot t}$, where $\lambda>0$ is a decay factor.
Since there is no global information about data streams, DenStream resorts to the idea of micro-clusters~\cite{10.5555/1315451.1315460} (\ie local stream information) to approximate the precise result in a streaming environment.
Each micro-cluster possesses three attributes: a weight $w$ indicating its commonness, which is determined based on the number and weight of points in it; a center $c$, which is the weighted center of the points in the micro-cluster; and a radius $r$, which is the weighted average of the distance from the points in the micro-cluster to the center.

Given the dynamic nature of evolving data streams, the roles of outliers and clusters are often exchanged.
Consequently, new clusters may emerge, and old ones gradually fade out.
To accommodate these continuous shifts, DenStream introduces two additional types of micro-clusters by setting different constraints on the weight, \ie Potential Micro Cluster (PMC) and Outlier Micro Cluster (OMC).
The PMC contains frequent and usual data points, while the OMC holds the points that could potentially be outliers or the seed of a new PMC (\ie previously unseen but new normal data points).
Thus, PMCs will have a larger weight than OMCs.
As time progresses, the number of data points within a cluster, along with their respective weights, will change.
This will in turn influence the overall weight of the cluster.
Based on such evolving weights, DenStream dynamically modifies the role of clusters.

Following the idea of DenStream, \name performs a continuous process to discover micro-clusters in streaming trace data and alter their role for trace sampling.
Before entering the online clustering process, \name first applies the DBSCAN algorithm~\cite{DBLP:conf/kdd/EsterKSX96} on the training data to generate the initial trace clusters.
In reality, the training data can be collected during the system's fault-free phases, which are easily obtainable since production services are mostly running in normal status.
The initial trace clusters represent the prevalent trace types within the system and act as a baseline to identify traces that deviate from these typical patterns.
When a new trace $T$ arrives, \name tries to merge it into existing micro-clusters as follows.

\begin{enumerate}
    \item At first, \name attempts to merge $T$ into its nearest PMC $C_p$, and calculates the new radius of $C_p$.
    If $C_p$'s new radius is below or equal to a predefined threshold $\epsilon$, which means the new trace $T$ is indeed similar to the members in $C_p$ (and thus a usual trace), the merge is successful.
    Next, the attributes of $C_p$ will be updated as follows.
    The weight $w$ is updated based on the formula $w_p^*=w_p\times 2^{-\lambda}+1$, where $w_p^*$ is the updated weight, $w_p$ is the previous weight, and $\lambda$ is the decay factor.
    The center $c$ is updated based on the formula $c_p^*=(c_p\times w_p\times 2^{-\lambda}+v_T)/w_p^*$, where $c_p^*$ is the updated center, $c_p$ is the previous center, and $v_T$ is the vector representation of trace $T$.
    Recall that the weight of a PMC is determined not only by the quantity of points it contains, but also by the individual weight of these points.
    The weight of $C_p$ increases with a larger number of points and with the recency of these points (as newer points have a larger weight).
    However, these two factors inversely affect the sampling probability of the traces in $C_p$.
    This is because we are more interested in the rare traces and those that haven't appeared recently.
    Therefore, we calculate the sampling probability $p_T$ for $T$, which is inversely proportional to the weight of $C_p$.
    We multiply the probability by the sampling budget $\mathcal{B}$ to meet the storage requirement:

    \begin{equation}
    \label{equ:sampling_probability}
        p_T = \mathcal{B}\times (1 - \frac{w_p^*}{\sum_{i=0}^{N_{pmc}} w^{(i)}})
    \end{equation}

    where $N_{pmc}$ is the total number of existing PMCs and $w^{(i)}$ is the weight of the $i$-th PMC.

    \item Else, if merging into the nearest PMC is unsuccessful, \name will try to merge $T$ into the nearest OMC $C_o$.
    Similarly, if the new radius of $C_o$ is below the predefined threshold $\epsilon$, the merge will be kept and the attributes of $C_o$ will be updated as in the first case.
    Then, \name checks whether the new weight of $C_o$ is above a noise threshold $\alpha$, which is the weight constraint for defining a PMC.
    If this is the case, it means $C_o$ has evolved into a PMC, \ie its trace pattern is deemed common.
    Thus, \name will switch its role accordingly.
    In this step, $T$ will be sampled no matter whether $C_o$'s role will be switched.
    This is because $T$ is dissimilar to the usual traces recorded in existing PMCs, but more similar to the traces in $C_o$ that are relative rare.

    \item Otherwise, if the previous two merging attempts fail, \name creates a new OMC containing only $T$.
    The weight and center of this new OMC are 1 and $T$, respectively.
    In this case, $T$ will be sampled, because it carries an unprecedented trace pattern that is not similar to any of the existing clusters.
    Note that the sampling of uncommon traces in this step (and step two) is not bound by the sampling budget.
    This is due to their rarity and inherent value, providing crucial insights into a system's edge-case behaviors.
\end{enumerate}

For each existing PMC, its weight will gradually decay if it fails to incorporate any new points.
If the weight is below the threshold $\alpha$, it means that the PMC has degraded back to an OMC, and it will be removed.
This is because the trace pattern represented by this PMC is considered expired, given no traces of this type have emerged for a certain period of time.
This could be attributed to system updates or the change of user behaviors, both of which could lead to the disappearance of certain types of traces.
In this way, we can maintain the latest trace patterns for online sampling.
To identify and remove the old PMCs, \name checks their weight periodically.
The time interval is calculated by the following formula:

\begin{equation}
    T_p = \lceil \frac{1}{\lambda} \textrm{log}(\frac{\alpha}{\alpha - 1}) \rceil
\end{equation}

In DenStream, the weight of existing OMCs also needs to be checked periodically.
OMCs falling below a certain weight limit are classified as outliers and subsequently removed.
However, this operation is not necessary in \name.
Our objective is to sample uncommon traces, which should include not just the outliers, but also the ones that represent normal system executions that have never appeared.
This is achieved by sampling traces associated with OMCs in both step two and three.
As the data stream proceeds, if an OMC indeed grows into a PMC, we will stop sampling from it.
Thus, there is no need for OMCs to undergo periodic checks or premature disposal, which saves computational overheads.
In this way, \name can identify diverse trace types while simultaneously preventing the over-sampling of recurring traces.


\subsection{Complexity Analysis}

\name performs two main operations to determine whether a trace should be sampled, \ie computing its sketch vector and merging it to micro-clusters.
In the worst case, the vector size of the trace is $|\mathcal{P}|$ (\ie it comprises all unique call paths) and the length of all its call paths is equal to $|p|_{max}$.
Then, computing its $L$-dimensional sketch vector requires a time complexity of $\mathcal{O}(|\mathcal{P}|\times L\times |p|_{max})$, which can be accelerated via matrix multiplication.
In reality, the complexity of a trace is likely to be much less than this worst-case scenario.
As for the merging part, the worst case is that the trace fails to join any of the existing micro-clusters.
The time complexity in this case is $\mathcal{O}((N_{pmc}+N_{omc})\times L)$, where $N_{omc}$ is the number of OMCs.
Since \name periodically fades out expired clusters, the total number of PMCs and OMCs remains small.
Therefore, we can conclude that the time complexity of \name is relatively low, rendering it a scalable trace sampler.
\section{Evaluation}
\label{sec:evaluation}

In this section, we present the evaluation of \name.
We first introduce the experimental settings, including the datasets, the metrics for evaluation, and the baseline methods.
Next, we demonstrate the experimental results, which include the effectiveness of trace sampling, the efficiency, and the sensitivity study of some important parameters.


The default parameter configurations in our experiments are as follows: the size of random numbers in each hash function $|p|_{max}=64$, the sketch length $L=100$, the threshold for declaring that a trace is close enough to a micro-cluster $\epsilon=0.01$.
All experiments are conducted on a Linux server with an Intel(R) Xeon(R) Gold 6226R CPU and 256GB RAM.
We repeat all experiments five times, which are averaged to yield the final results.

\subsection{Experimental Settings}

\subsubsection{Datasets}

We collect traces from two open-source benchmark microservice applications, \ie Train Ticket~\cite{DBLP:journals/tse/ZhouPXSJLD21} and Online Boutique~\cite{onlineboutique}, which have been widely used in previous studies~\cite{DBLP:conf/sigsoft/YuCLCLZ23,DBLP:conf/sigcomm/LiCL19,DBLP:conf/sosp/LoffP0M023}.
Train Ticket is a railway ticketing system with 41 microservices, where users can search, book, pay, and cancel train tickets.
Online Boutique is a web-based e-commerce app with 11 microservices, where users can browse items, add them to the cart, and make a payment.
These benchmarks are implemented in different programming languages such as Java, Go, Node.js, Python, etc.
We follow~\cite{DBLP:conf/sigsoft/YuCLCLZ23} to deploy these applications and inject faults to generate abnormal traces that correspond to edge-case system behaviors.
For each application, we synthesize workloads via the load testing tool Locust~\cite{locust} to simulate user requests.
We label two types of uncommon traces that are deemed necessary for sampling.
The first is the abnormal traces generated during fault injection, and the second is those corresponding to new service operations that are not previously observed in the training data.

Besides the benchmark applications, we also evaluate \name on a production dataset, which comprises hundreds of thousands of traces from large-scale cloud service systems.
These services run within containers that are directly managed by Kubernetes~\cite{kubernetes}.
The traces to be sampled are labeled by on-site engineers based on their domain knowledge.
Table~\ref{tab:data_statistics} summarizes the statistics of the three trace datasets.
\textit{\#Span} denotes the number of different span types in the dataset.
We split each dataset into training and testing sets, whose size is given by \textit{\#Train} and \textit{\#Test}.
In particular, as our goal is to design a streaming trace sampler, we keep a low ratio for the training set in order to effectively evaluate \name in real-world conditions.
The traces in the testing set are gradually fed into the model based on their timestamps to simulate an online scenario.
Finally, \textit{\#Label} gives the number of labeled traces, which we aim to sample.

\begin{table}[t]
\begin{center}
\caption{Dataset Statistics}
\label{tab:data_statistics}
\normalsize
 \begin{tabular}{c|c|c|c|c}
  \toprule
  
  \textbf{Dataset} & \textbf{\#Span} & \textbf{\#Train} & \textbf{\#Test} & \textbf{\#Label}\\
  
  \midrule
  \midrule

  Train Ticket & 201 & 653 & 31,814 & 624 \\
  Online Boutique & 43 & 1,024 & 78,931 & 225 \\
  Industry & 1,308 & 11,847 & 497,439 & 2,453 \\

  \bottomrule
 \end{tabular}
\end{center}
\end{table}

\subsubsection{Evaluation Metrics}
Intuitively, there are two essential aspects to gauge the effectiveness of a trace sampling algorithm.
The first is whether the sampler can successfully capture the useful traces.
We refer to this metric as \textit{Coverage}, which can be calculated as the ratio of labeled traces sampled by the algorithm to the total number of labeled traces.
The second is at what \textit{Sampling Rate} the coverage is accomplished.
This metric can be calculated as the ratio of the number of traces sampled to the total number of traces available in the testing set.
An ideal trace sampler should be able to capture a broad spectrum of useful traces (\ie a high Coverage), while maintaining resource efficiency (\ie a low Sampling Rate).
In reality, missing important traces could lead to their permanent loss, particularly in online scenarios.
As a result, engineers may lose valuable insights into the functioning and performance of the system.
Thus, the Coverage should generally be prioritized over the Sampling Rate.

\subsubsection{Baseline Methods}
\label{sec:baselines}

The following methods are selected for a comparative evaluation of \name.
The default sampling budget for all methods is set as 1\%, except for Sieve which does not require this parameter.

\begin{itemize}
    \item \textit{Uniform}: This strategy uniformly samples a trace subset from the testing set, without considering any specific characteristics of the traces.
    It is the default sampling mechanism used in many distributed tracing tools, \ie the head sampling.

    \item \textit{Sieve}: Sieve~\cite{DBLP:conf/icws/HuangCYCZ21} is an online trace sampler that leverages an attention mechanism to bias sampling toward uncommon traces.
    It uses the technique of Robust Random Cut Forest (RRCF), which is a variant of the Isolation Forest, to calculate an attention score for traces.
    Uncommon traces tend to have a shallower depth from the root to leaf, and thus will receive more attention (\ie a higher probability) for sampling.
    New dimensions are integrated into the model by appending them to the feature vectors, and the tree structure is subsequently modified to align with these additions.

    \item \textit{Sifter}: Sifter~\cite{DBLP:conf/cloud/Las-CasasPAM19} captures edge-case traces by learning an unbiased, low-dimensional model based on fixed-length trace sub-paths.
    Such a model is able to approximate the system's common-case behaviors.
    Thus, by measuring the reconstruction loss of an incoming trace, the sampling decisions can be made toward traces that lead to a large loss.
    Particularly, Sifter only considers the structural feature of traces, and is thus insensitive to temporal deviations.

    \item \textit{SampleHST}: Based on the Bag-of-Words (BoW) representation of traces, SampleHST~\cite{DBLP:conf/noms/GiasGSPOC23} calculates a distribution of the mass values obtained from a forest of tree-based classifier, \ie Half Space Trees (HSTs).
    The mass distribution derived from the HSTs is then utilized to cluster the traces online, leveraging a variant of the mean-shift algorithm.
    A trace is more likely to be sampled if it is associated with a low-mass-value cluster.
    Similar to Sifter, the BoW representation makes SampleHST unable to consider the temporal feature of traces.

    \item \textit{Perch}: Perch~\cite{DBLP:conf/cloud/Las-CasasMGF18} represents traces based on various graph-based features, \eg occurrence count embeddings based on user-specified events.
    Then, a hierarchical clustering algorithm named PERCH is utilized to groups traces.
    Representative traces are then evenly selected from each trace group.
    
\end{itemize}

\subsection{Experimental Results}

\subsubsection{Trace Sampling}

\begin{table*}[t]
\begin{center}
\caption{Experimental Results of Trace Sampling}
\label{tab:exp_trace_sampling}
\normalsize
 \begin{tabular}{c|c|c|c|c|c|c}
  \toprule
  & \multicolumn{2}{c|}{Train Ticket} & \multicolumn{2}{c|}{Online Boutique} & \multicolumn{2}{c}{Industry} \\
  
  \textbf{Method}  & \textbf{Coverage} &  \textbf{Sampling Rate} & \textbf{Coverage} & \textbf{Sampling Rate} & \textbf{Coverage} & \textbf{Sampling Rate}\\
  
  \midrule
  \midrule

Uniform   & 1.1\%  & \textbf{1.00\%} & 1.1\%  & 1.00\% & 1.0\% & 1.00\% \\
Sieve     & 71.8\% & 6.24\% & 82.2\% & 3.91\% & 61.1\% & 3.29\% \\
Sifter    & 50.6\% & 1.28\% & 42.2\% & 1.13\% & 37.3\% & 1.16\%\\
SampleHST & 46.2\% & 1.19\% & 38.2\% & 1.07\% & 34.5\% & 1.10\% \\
Perch     & 42.9\% & \textbf{1.00\%} & 35.1\% & \textbf{0.98\%} & 32.7\% & 0.95\%\\
\midrule
\name     & \textbf{98.7\%} & 2.34\% & \textbf{100\%}  & 1.35\% & \textbf{93.6\%} & \textbf{0.83\%} \\
  \bottomrule
 \end{tabular} 
\end{center}
\end{table*}

Table~\ref{tab:exp_trace_sampling} shows the performance of trace sampling of different methods on three datasets, including both the coverage and sampling rate.
We can see that \name achieves the best coverage score on all datasets.
The score on the Industry dataset is comparatively lower due to its scale and complexity.
Sieve ranks second in terms of the coverage performance.
However, as it lacks explicit consideration of the sampling budget, it has the highest sampling rate in all cases.
On the Train Ticket and Online Boutique datasets, \name also has a relatively high sampling rate.
This is because in our design (Section~\ref{sec:trace_sampling}), rare traces are deemed necessary for sampling regardless of the sampling budget, as they provide essential insights into a system's edge-case behaviors.
Nevertheless, \name demonstrates the best sampling rate on the Industry dataset.
We can impose a harder budget constraint to further lower the sampling rate, \ie an even smaller $\mathcal{B}$ in Equation (\ref{equ:sampling_probability}).
It will only decreases the sampling of traces that are relatively common.
Uniform sampling achieves the best sampling rate (\ie 1\%), but with the worst coverage, which is also around 1\%.
The remaining three methods also present a good sampling rate.
This can be attributed to their relatively stringent budget constraints, particularly in the case of Perch.
However, they fall short in terms of coverage scores.
An important reason is that they only consider the structural features of traces.
This observation underscores the significance of temporal features.

In particular, certain approaches \eg~\cite{DBLP:conf/cloud/Las-CasasMGF18}, perform representative sampling, where only a few traces are selected from each cluster of similar traces.
However, in real-world situations, when performance issues arise, it becomes essential to capture the entire spectrum of edge-case traces regardless of their similarity.
This not only provides a more comprehensive understanding of the issue at hand (\eg the blast radius~\cite{DBLP:conf/sigsoft/ChenKLZZXZYSXDG20}) but also ensures that the solutions are robust and effective.
\name mitigates this problem by continuously collecting traces that belong to OMCs, while also allows the dynamic transformation of a trace cluster's role to avoid over-sampling.

\begin{figure}
    \centering
    \includegraphics[width=0.9\linewidth]{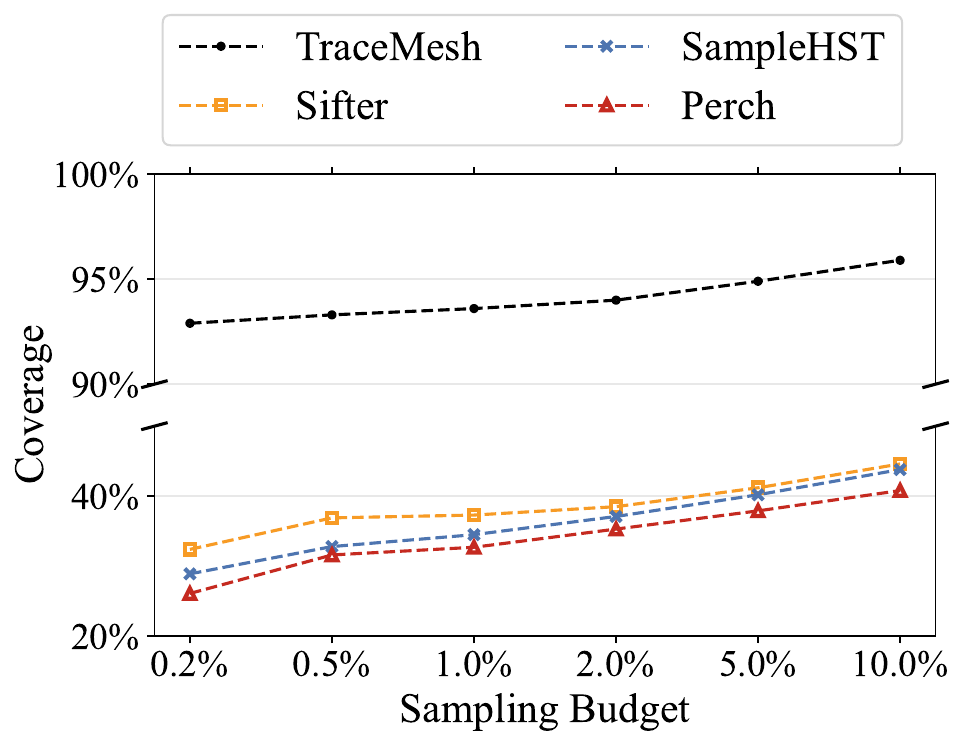}
    \caption{Coverage with different sampling budgets}
    \label{fig:sampling_budget}
\end{figure}

We further evaluate the coverage of different methods with various sampling budgets, as shown in Figure~\ref{fig:sampling_budget}.
We exclude Sieve due to its lack of a budget-control design.
As Industry is the most challenging dataset, we only show the results derived from it.
Other datasets demonstrate similar results.
Clearly, as the sampling budget grows, all methods achieve better coverage scores.
In all situations, \name outperforms the baselines by a large margin, demonstrating both superior performance and stability.
This is because \name can automatically identify the most important traces for sampling, even when the sampling budget is exceeded.
It is worth noting that, edge cases are rare in production systems, and the storage budget is typically sufficient.
The key issue lies in how to accurately identify such valuable traces.

\subsubsection{Efficiency}

\begin{figure*}
    \centering
    \includegraphics[width=0.9\linewidth]{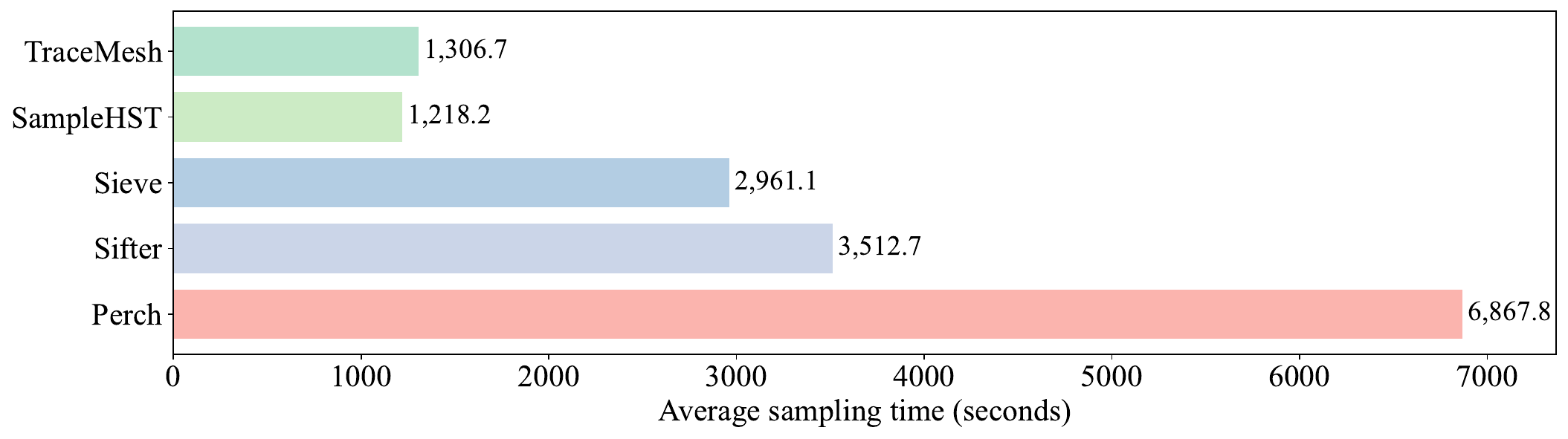}
    \caption{Efficiency on the Industry dataset}
    \label{fig:efficiency}
\end{figure*}

In production cloud service systems, the sheer volume and complexity of traces pose significant challenges on the efficiency of trace sampling.
Thus, we evaluate different methods in this aspect, as shown in Figure~\ref{fig:efficiency}.
We only present results for the Industrial dataset, which has a significantly higher feature dimension compared to other datasets.
The efficiency is quantified by the time taken by each method to complete the sampling process for the Industry dataset.
The performance of different methods can be divided into three tiers.
The first tier includes SampleHST and our method, taking the minimum time, \ie 1,218.2 and 1,306.7 seconds respectively.
SampleHST employs HSTs to compute the mass distribution of traces, making it a lightweight model with a constant amortized time complexity.
The performance of \name can be attributed to its efficient sketching mechanism and evolving clustering design.
Particularly, in Figure~\ref{fig:trace_sketching}, for the zero entries of the trace vector (\ie the trace does not have the corresponding call paths), the sketching process can be omitted, thereby further enhancing its efficiency.
Sieve and Sifter belong to the next tier, which require a considerably longer duration of 2,961.1 and 3,512.7 seconds respectively.
This indicates that the second tier operates nearly three times slower than the first for trace sampling. 
In online scenario, Sieve needs to adjust its tree structures to accommodate new feature dimensions.
This process is expensive, especially when dimension expansion happens frequently.
Sifter involves embedding calculation, which can be time-consuming.
Perch alone constitutes the third tier, which demands the maximum time, \ie 6,867.8 seconds.
Similar to Sieve, Perch requires continuous tree structure modifications, a process that becomes increasingly slower with the addition of more traces.

\begin{figure}
    \centering
    \includegraphics[width=0.9\linewidth]{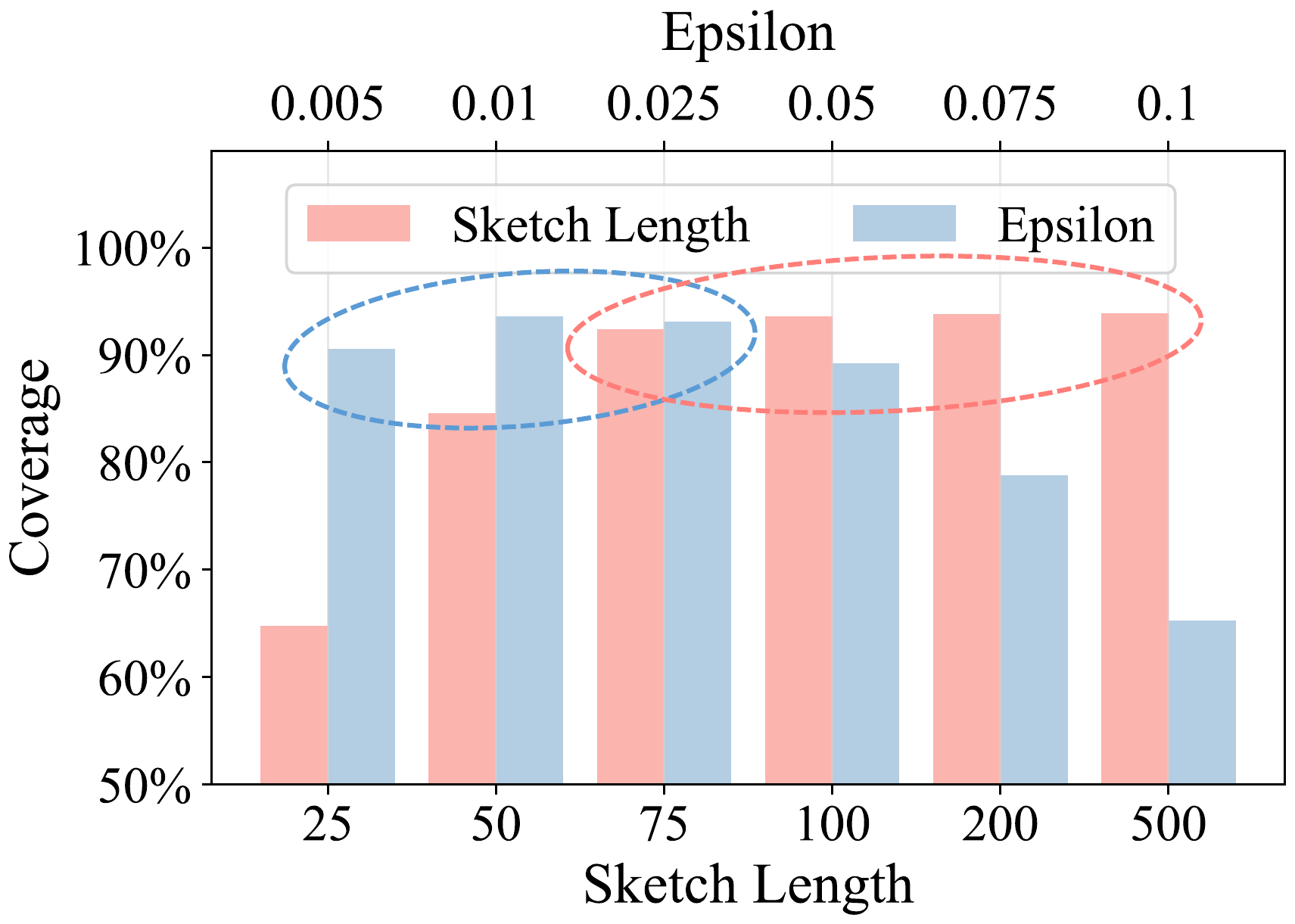}
    \caption{Parameter sensitivity analysis}
    \label{fig:sensitivity}
\end{figure}

\subsubsection{Sensitivity Analysis}

The sensitivity to parameter configurations is paramount for a method to consistently deliver stable performance.
We conduct a sensitivity analysis of \name with regard to two critical parameters, \ie the sketch length ($L$) and epsilon ($\epsilon$), as shown in Figure~\ref{fig:sensitivity}.
\name employs LSH to efficiently measure the cosine similarity between two trace vectors in an $L$-dimensional space.
A large $L$ can reduce the error of cosine distance approximation but at the cost of increased computational overhead.
The default setting of $L$ in our experiments is 100.
After trying a larger spectrum of settings, \ie from 25 to 500, we can see that \name's performance converges when $L$ is larger than 75.
Another parameter, $\epsilon$, is used to determine whether a new trace belongs to a micro-cluster (Section~\ref{sec:trace_sampling}).
This is significant because it impacts the evolution process of micro-clusters, such as when an OMC evolves into a PMC.
It can be observed that \name yields the best performance when $\epsilon$ lies within the range of $[0.005, 0.025]$.
In light of our findings, \name demonstrates a good ability to maintain consistent performance across various parameter settings.

\section{Related Work}
\label{sec:related_work}

Distributed tracing has emerged as a crucial tool for understanding and optimizing complex service-oriented architectures.
It provides an invaluable mechanism for tracking requests as they navigate through the intertwined services of a distributed system, enabling developers to identify bottlenecks, spot inefficiencies, and troubleshoot performance issues.
This area of study has seen significant contributions from various efforts.
X-Trace~\cite{DBLP:conf/nsdi/FonsecaPKSS07} aims to provide a comprehensive view of service behavior across multiple system layers.
Dapper~\cite{sigelman2010dapper} is a distributed system tracing infrastructure developed by Google.
Its design goals include low overhead, application-level transparency, and ubiquitous deployment across extensive, large-scale systems.
Pivot Tracing~\cite{DBLP:conf/sosp/MaceRF15} considers the \textit{happened-before} relations between events during dynamic instrumentation.
This provides users with the ability to define arbitrary metrics for mining useful information about root causes at runtime.
Canopy~\cite{DBLP:conf/sosp/KaldorMBGKOOSSV17}, developed by Facebook, is an end-to-end performance tracing infrastructure.
It captures performance data with causal relationships across various platforms including browsers, mobile applications, and backend services.
This aids engineers in querying and analyzing performance data in real-time.
DeepFlow~\cite{DBLP:conf/sigcomm/ShenZXSLSZWY0XL23} is a non-intrusive distributed tracing framework for troubleshooting microservices.
It provides out-of-the-box tracing via a network-centric tracing plane and implicit context propagation.
In recent years, open-source tracing frameworks like Jaeger~\cite{jaeger} and Zipkin~\cite{zipkin} have seen widespread adoption in practical applications.

Besides the infrastructure for generating distributed traces, trace data are widely used in various system reliability assurance tasks.
The empirical study in~\cite{DBLP:journals/tse/ZhouPXSJLD21} shows that the current industrial practices of microservice debugging can be improved through the application of appropriate tracing and visualization techniques.
MEPFL~\cite{DBLP:conf/sigsoft/Zhou0X0JLXH19} leverages system trace logs in the production environment to predict latent errors, faulty microservices, and fault types at runtime.
Groot~\cite{DBLP:conf/kbse/WangWJHWKX21} builds a real-time causality graph to facilitate root cause analysis in microservices, which allows adaptive customization of link construction rules to incorporate domain knowledge.
TraceStream~\cite{DBLP:conf/issre/ZhouZPYLLZZD23} identifies potential clusters of anomalous traces within evolving trace data and employs spectral analysis to pinpoint the specific 
anomalous services.
Some work~\cite{DBLP:conf/kbse/ChenLSZWLYL21,DBLP:conf/icse/LeeYCSL23,DBLP:conf/sigsoft/YuCLCLZ23} utilizes a multi-modal approach, which integrates traces with logs and metrics to provide more comprehensive information about system status for microservice troubleshooting.
Traces also serve a crucial role in in analyzing system dependencies~\cite{DBLP:conf/cloud/LuoXLYXZDH021}, critical paths~\cite{DBLP:conf/usenix/0002RRPSC22}, resource characterization~\cite{DBLP:conf/icpp/WangLWJCWDXHYZ22,DBLP:conf/kbse/LiuJGHCFYYL23}, and microservice architecture~\cite{DBLP:conf/usenix/HuyeSS23,DBLP:conf/sigsoft/0001ZZIGC22}.
This enables developers to identify bottlenecks, spot inefficiencies, and subsequently optimize the overall system performance.

With the dramatic increase in trace volume in production systems, trace sampling has become an essential technique to manage data overload and maintain system efficiency.
Representative approaches are introduced in Section~\ref{sec:baselines}, which leverage a variety of methods, including tree-based models~\cite{DBLP:conf/icws/HuangCYCZ21,DBLP:conf/noms/GiasGSPOC23}, clustering algorithms~\cite{DBLP:conf/cloud/Las-CasasMGF18}, and neural language techniques~\cite{DBLP:conf/cloud/Las-CasasPAM19}.
STEAM~\cite{DBLP:conf/sigsoft/HeFLZ0LR023} preserves system observability by sampling mutually dissimilar traces.
It employs Graph Neural Networks (GNN) for trace representation~\cite{DBLP:conf/sigsoft/Zhang0ZSYCY22}, and requires human labeling to incorporate domain knowledge.
Hindsight~\cite{DBLP:conf/nsdi/ZhangXAVM23} introduces the idea of \textit{retroactive sampling}.
Instead of eagerly ingesting and processing traces, it lazily retrieves trace data only after symptoms of a problem are detected.
Hindsight is based on the observation that the generation of traces at nodes is not expensive, but the ingestion of trace data is resource-intensive.
Thus, tail sampling strategy can be applied to achieve accurate trace sampling without occurring too much runtime overhead (\ie before the processing of trace data).

While progress has been made, existing learning-based approaches for trace sampling lack consideration of practical challenges associated with their deployment in production environments.
Specifically, the large volume of trace data and feature space demand a highly efficient solution.
The diversity of traces also requires fast ability to accommodate new dimensions.
Unlike these works, our design emphasizes operational efficiency and flexibility.
It not only handles large volumes of trace data with a vast feature space, but also quickly adapts to the emergence of unprecedented trace features.
\section{Conclusion}
\label{sec:conclusion}

In this paper, we propose \name, a scalable and streaming trace sampler.
It addresses the practical challenges of deploying trace sampling techniques in production systems, which have not been adequately addressed in existing work.
The scale and complexity of modern service systems render trace data highly diverse and dynamic.
The unpredictable system updates and user behaviors further compound the situation.
Thus, a practical sampler should be able to handle the vast feature space of traces and quickly generalize to previously unknown features within the trace stream.
To this end, we first perform trace vector encoding by considering both the structural and temporal features.
Next, we employ streaming Locality-Sensitivity Hashing (LSH) technique to enable efficient trace similarity mining by projecting them into a low-dimensional space.
In this process, new features can be seamlessly incorporated without changing the input dimensionality.
Finally, given a sampling budget, we identify and capture uncommon traces through evolving clustering.
We dynamically adjust the sampling decision to prevent accumulating redundant information.
We have evaluated \name using trace data collected from both open-source microservice benchmarks and production service systems.
Experimental results highlight its potential to pave the way for efficient system monitoring.
\section*{Acknowledgement}

This work is supported by the National Natural Science Foundation of China (No. 62202511) and the Research Grants Council of the Hong Kong Special Administrative Region, China (No. CUHK 14206921 of the General Research Fund).


\balance
\bibliographystyle{IEEEtran}
\bibliography{bibliography}
\balance

\end{document}